\documentclass[12pt]{iopart}

\usepackage{graphicx}
\usepackage{xspace}

\def\cref#1{Chapt.\,\ref{#1}}
\def\Cref#1{Chapter~\ref{#1}}

\def\fref#1{Fig.\,\ref{#1}}
\def\ffref#1{Figs.\,\ref{#1}}

\def\rref#1{Ref.\,\cite{#1}}

\def\lleft{\textit{left}}
\def\rright{\textit{right}}

\def\ttop{\textit{top}}
\def\bbottom{\textit{bottom}}

\def\1{\footnotemark[1]}
\def\and{\& }

\def\deg{$^\circ$\xspace}

\def\gcm2{g/cm$^2$\xspace}

\def\lg{{\rm lg}\xspace}

\begin{document}

\title[EPOS tests with air shower data]
      {Test of the hadronic interaction model EPOS with air shower data}

\author{
W.D.~Apel$^1$, 
J.C.~Arteaga$^{1}$\footnote{now at Institute of Physics and Mathematics,
Universidad Michoacana, Morelia, Mexico},
F.~Badea$^{1}$, 
K.~Bekk$^1$, 
M.~Bertaina$^2$,
J.~Bl\"umer$^{1,3}$,
H.~Bozdog$^1$,
I.M.~Brancus$^4$,
M.~Br\"uggemann$^5$,
P.~Buchholz$^5$,
E.~Cantoni$^{3,7}$,
A.~Chiavassa$^2$,
F.~Cossavella$^3$, 
K.~Daumiller$^1$, 
V.~de Souza$^3$\footnote{now at Universidade de S{\~a}o Paulo, Instituto de
Fisica de S{\~a}o Carlos, Brasil}, 
F.~Di~Pierro$^2$,
P.~Doll$^1$, 
R.~Engel$^1$,
J.~Engler$^1$, 
M.~Finger$^1$, 
D.~Fuhrmann$^6$,
P.L.~Ghia$^7$,
H.J.~Gils$^1$,
R.~Glasstetter$^6$, 
C.~Grupen$^5$,
A.~Haungs$^1$, 
D.~Heck$^1$, 
J.R.~H\"orandel$^{3}$\footnote{corresponding author
     email:j.horandel@astro.ru.nl, now at Department of Astrophysics, Radboud
     University Nijmegen, The Netherlands}, 
T.~Huege$^1$, 
P.G.~Isar$^1$, 
K.-H.~Kampert$^6$,
D.~Kang$^3$,
D.~Kickelbick$^5$,
H.O.~Klages$^1$, 
Y.~Kolotaev$^5$,
P.~Luczak$^8$, 
H.J.~Mathes$^1$, 
H.J.~Mayer$^1$, 
J.~Milke$^1$, 
B.~Mitrica$^4$,
C.~Morello$^7$,
G.~Navarra$^2$,
S.~Nehls$^1$,
J.~Oehlschl\"ager$^1$, 
S.~Ostapchenko$^{1}$\footnote{now at Norwegian University, Trondheim, Norway},
S.~Over$^5$,
M.~Petcu$^4$, 
T.~Pierog$^1$, 
H.~Rebel$^1$, 
M.~Roth$^1$, 
H.~Schieler$^1$, 
F.~Schr\"oder$^1$,
O.~Sima$^9$, 
M.~St\"umpert$^3$, 
G.~Toma$^4$, 
G.C.~Trinchero$^7$,
H.~Ulrich$^1$,
J.~van~Buren$^1$,
W.~Walkowiak$^5$,
A.~Weindl$^1$,
J.~Wochele$^1$, 
M.~Wommer$^1$,
J.~Zabierowski$^8$
}

\address{
$^{1}$ Institut\ f\"ur Kernphysik, Forschungszentrum Karlsruhe,
76021~Karlsruhe, Germany\\
$^{2}$ Dipartimento di Fisica Generale dell'Universit{\`a},
10125 Torino, Italy\\
$^{3}$ Institut f\"ur Experimentelle Kernphysik,
Universit\"at Karlsruhe, 76021 Karlsruhe, Germany,\\
$^{4}$ National Institute of Physics and Nuclear Engineering,
7690~Bucharest, Romania\\
$^{5}$ Fachbereich Physik, Universit\"at Siegen, 57068 Siegen, 
Germany \\
$^{6}$ Fachbereich Physik, Universit\"at Wuppertal, 42097
Wuppertal, Germany \\
$^{7}$ Istituto di Fisica dello Spazio Interplanetario, INAF, 
10133 Torino, Italy \\
$^{8}$ Soltan Institute for Nuclear Studies, 90950~Lodz, 
Poland\\
$^{9}$ Department of Physics, University of Bucharest, 
76900~Bucharest, Romania\\
}

\begin{abstract}
Predictions of the hadronic interaction model EPOS~1.61 as implemented in the
air shower simulation program CORSIKA are compared to observations with the
KASCADE experiment. The investigations reveal that the predictions of EPOS are
not compatible with KASCADE measurements.  The discrepancies seen are most
likely due to use of a set of inelastic hadronic cross sections that are too
high.
\end{abstract}

\vspace{2pc}
\submitto{\JPG}

\section{Introduction}

When high-energy cosmic rays penetrate the Earth's atmosphere they initiate
cascades of secondary particles --- the extensive air showers.  Objective of
air shower detectors is to derive information about the shower inducing primary
particle from the registered secondary particles.  Addressing astrophysical
questions with air-shower data necessitates the understanding of high-energy
interactions in the atmosphere.  Or, in reversion, the interpretation of
properties of primary radiation derived from air-shower measurements depends on
the understanding of the complex processes during the development of air
showers.  
In the last decade significant progress has been achieved in the interpretation
of air shower data and main properties of the primary cosmic radiation have
been measured. At energies around $10^{6}$~GeV the mass composition of cosmic
rays has been investigated and energy spectra for groups of elements could be
derived \cite{ulrichapp,eastopspec}. It could be shown that the knee in the
all-particle energy spectrum at about $4\cdot10^6$~GeV is caused by a cut-off
in the energy spectra of the light elements (protons and helium).
Despite of this progress, detailed investigations indicate inconsistencies in
the interpretation of air shower data
\cite{ulrichapp,rothnn,chicagoknee,pg,wq,wwtestjpg,jensjpg}.
Thus, one of the goals of KASCADE (Karlsruhe Shower Core and Array DEtector) is
to investigate high-energy interactions in the atmosphere and to improve
contemporary models to describe such processes.

For air shower interpretation the understanding of multi-particle production in
hadronic interactions with a small momentum transfer is essential
\cite{engelpylos}.  Due to the energy dependence of the coupling
constant $\alpha_s$ soft interactions cannot be calculated within QCD using
perturbation theory. Instead, phenomenological approaches have been introduced
in different models. These models are the main source of uncertainties in
simulation codes to calculate the development of extensive air showers, such as
the program CORSIKA \cite{corsika}.  Several codes to describe hadronic
interactions at low energies ($E<200$~GeV; e.g.\ GHEISHA \cite{gheisha} and
FLUKA \cite{flukacern,flukaCHEN}) as well as high energies (e.g.\ DPMJET
\cite{dpmjet}, QGSJET \cite{qgsjet,qgsII,qgsjetII}, SIBYLL \cite{sibyll21}, and
EPOS \cite{epos,epos2}) have been embedded in CORSIKA.  

The test of interaction models necessitates detailed measurements of several
shower components. The KASCADE experiment \cite{kascadenim} with its
multi-detector set-up, registering simultaneously the electromagnetic, muonic,
and hadronic shower components is particularly suited for such investigations.
The information derived on properties of high-energy interactions from air
shower observations is complementary to measurements at accelerator experiments
since different kinematical and energetic regions are probed. 

In previous investigations \cite{wwtestjpg,jensjpg} the models QGSJET versions
98 and 01 \cite{qgsjet}, VENUS \cite{venus}, SIBYLL versions 1.6
\cite{sibyll16} and 2.1 \cite{sibyll21}, DPMJET \cite{dpmjet}, and {\sc neXus}
\cite{nexus} have been studied.  The analyses presented in this article focus
on the interaction model EPOS, version 1.61. This model is a recent development,
historically emerging from the VENUS and {\sc neXus} codes.

EPOS is a consistent quantum mechanical multiple scattering approach based on
partons and strings, where cross sections and the particle production are
calculated consistently, taking into account energy conservation in both cases
(unlike other models where energy conservation is not considered for cross
section calculations~\cite{hladik2001}).  A special feature is the explicit
treatment of projectile and target remnants, leading to a better description of
baryon and antibaryon production than in other models used for cosmic-ray
analysis.  Motivated by the data obtained by the RHIC experiments, nuclear
effects related to Cronin transverse momentum broadening, parton saturation,
and screening have been introduced into EPOS. Furthermore, unlike other models,
high density effects  leading to collective behavior in heavy ion collisions
(or lighter systems) are also taken into account. Since this model is applied
to accelerator physics, many data are considered which are not a priori linked
to cosmic rays and air showers. That is may be the largest difference to all
other hadronic models used to simulate air showers.

\section{Experimental set-up}

\subsection{The apparatus}

The experiment KASCADE, located on site of the Forschungszentrum Karlsruhe, 110
m a.s.l., consists of several detector systems. A description of the
performance of the experiment can be found elsewhere \cite{kascadenim}. A $200
\times 200$~m$^2$ array of 252 detector stations, equipped with scintillation
counters, measures the electromagnetic and, below a lead/iron shielding, the
muonic parts of air showers. In its center, an iron sampling calorimeter of $16
\times 20$~m$^2$ area detects hadronic particles. The calorimeter is equipped
with 11 000 warm-liquid ionization chambers arranged in nine layers. Due to its
fine segmentation ($25\times25$~cm$^2$), energy, position, and angle of
incidence can be measured for individual hadrons. A detailed description of the
calorimeter and its performance can be found in \cite{kalonim}, it has been
calibrated with a test beam at the SPS at CERN up to 350~GeV particle energy
\cite{kalocern}.

\subsection{Observables and event selection}

The position of the shower axis and the angle of incidence of a cascade are
reconstructed by the array detectors. The total numbers of electrons $N_e$ and
muons $N_\mu$ are determined by integrating their lateral distributions. In
case of muons, the ‘truncated muon number’ $N_\mu^{tr}$ is used for experimental
reasons. It is the number of muons integrated in the distance range $40-200$~m
from the shower axis. For a detailed description of the reconstruction
algorithms see \cite{kascadelateral}.
The position of the shower axis is reconstructed with an accuracy better than
2~m and the angle of incidence better than 0.5\deg.  

The hadrons in the calorimeter are reconstructed by a pattern recognition
algorithm, optimized to recognize as many hadrons in a shower core as possible.
Details can be found in \cite{kascadelateral}.  Hadrons of equal energy can
still be separated with a probability of 50\% at a distance of 40~cm. The
reconstruction efficiency rises from 70\% at 50~GeV to nearly 100\% at 100~GeV.
The energy resolution improves from 30\% at 50~GeV to 15\% at $10^4$~GeV.  The
hadron number $N_h$ and hadronic energy sum $\sum E_h$ are determined by the
sum over all hadrons in a distance up to 10~m from the shower axis. A
correction for the missing area beyond the boundaries of the calorimeter is
applied. In the following, $N_h$ and $\sum E_h$ are given for a threshold of
100~GeV, but also hadronic shower sizes for higher thresholds up to 500~GeV
have been investigated. The observable $\sum E_h$ includes also energy of
hadrons which could not be reconstructed independently, because they are too
close to each other. It shows up in the simulated and experimental data in the
same manner.  

To be accepted for the analysis, an air shower has to fulfill several
requirements: at least one hadron has been reconstructed in the calorimeter
with an energy larger than 50~GeV, the shower axis is located inside the
calorimeter, the electromagnetic shower size $N_e$ is larger than $10^4$, the
truncated muon number $N_\mu^{tr}$ is larger than $10^3$, i.e.\ the primary
energy is greater than about $3\cdot10^5$~GeV, and the reconstructed zenith
angle is smaller than 30\deg.
For \ffref{qgs2} and \ref{nenmplane} different selection criteria have been
applied \cite{ulrichapp}. Namely: the reconstructed shower axis has to be
within 91~m from the center of the array, the age parameter $s$, obtained
through a fit of a NKG function to the lateral distribution of the
electromagnetic component has to be in the interval $0.2<s<2.1$, and only
showers with $\lg N_e\ge4.8$, $\lg N_\mu^{tr}\ge3.6$, as well as a zenith angle
$<18^\circ$  are considered.

\subsection{Simulations}

The shower simulations were performed using CORSIKA.  Hadronic interactions at
low energies were modeled using the FLUKA code \cite{flukacern,flukaCHEN}.
High-energy interactions were treated with EPOS~1.61 \cite{epos,epos2}
($E>80$~GeV) as well as QGSJET~01 \cite{qgsjet} ($E>200$~GeV).  The latter has
been chosen for reference in order to compare the results discussed in the
present article to previous publications \cite{wwtestjpg,jensjpg}.  Showers
initiated by primary protons and iron nuclei have been simulated.  The
simulations covered the energy range $10^{5}-10^{8}$~GeV with zenith angles in
the interval $0^\circ - 32^\circ$. The spectral index in the simulations was
$-2.0$. For the analysis it is converted to a slope of $-2.7$ below and $-3.1$
above the knee with a rigidity dependent knee position ($3\cdot10^6$~GeV for
protons). \footnote{Again, \ffref{qgs2} and \ref{nenmplane} have been treated
differently, see \rref{ulrichapp}.} The positions of the shower axes are
distributed uniformly over an area exceeding the calorimeter surface by 2~m on
each side. In order to determine the signals in the individual detectors, all
secondary particles at ground level are passed through a detector simulation
program using the GEANT package \cite{geant}.  In this way, the instrumental
response is taken into account and the simulated events are analyzed by the
same code as the experimental data, an important aspect to avoid biases by
pattern recognition and reconstruction algorithms.  

\begin{figure}
 \includegraphics[width=0.49\textwidth]{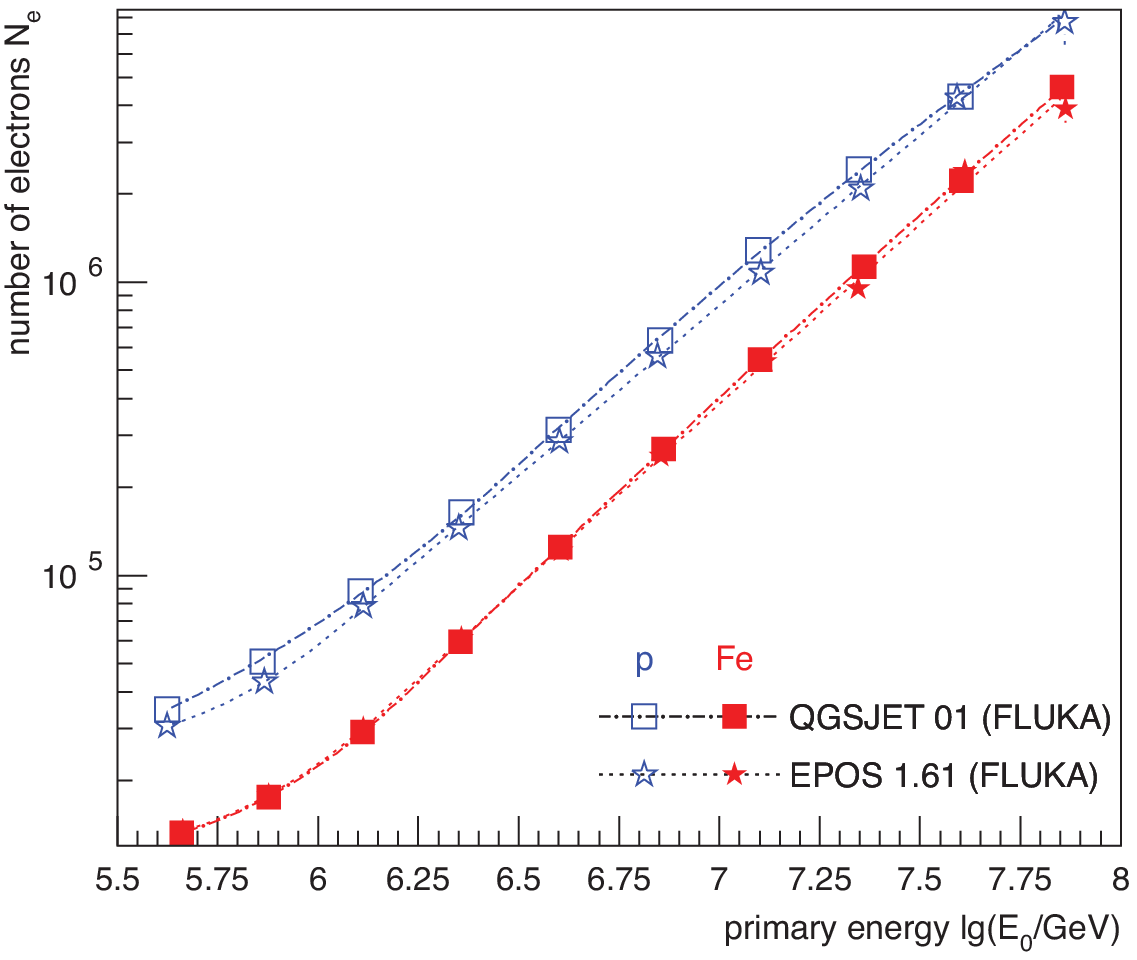}\hspace*{\fill}
 \includegraphics[width=0.49\textwidth]{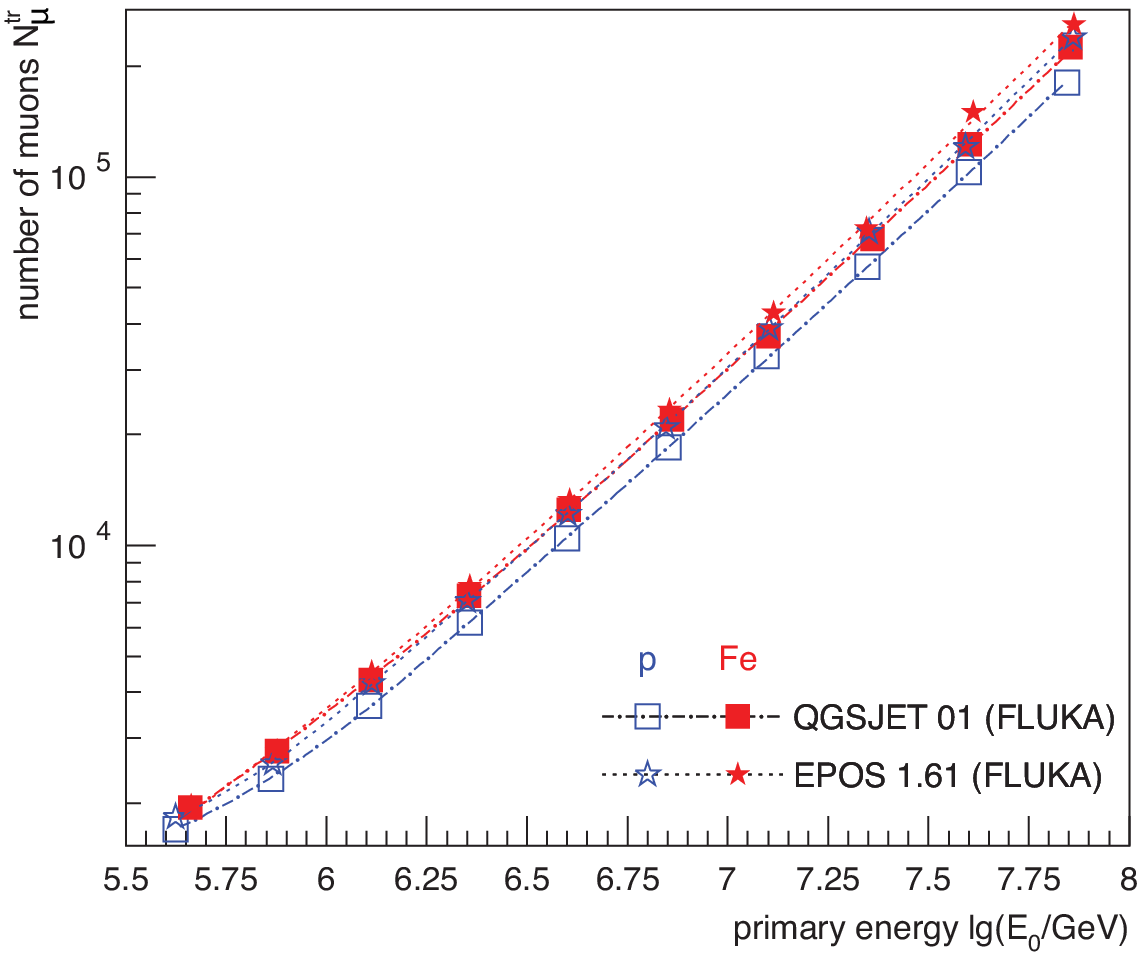}
 \caption{Number of electrons (\lleft) and number of muons (\rright) as
	  function of shower energy for proton and iron induced showers as
          predicted by the hadronic interaction models EPOS and QGSJET~01.}
 \label{nenme}	  
\end{figure}

The average primary energy belonging to a simulated and reconstructed number of
electrons and muons is given in \fref{nenme}. The left panel demonstrates the
$N_e$ dependence on the primary mass. The lines through the points are drawn to
guide the eye and represent five parameter fits. As in all figures errors of
the mean values are plotted. But, in most cases, the error bars are smaller
than the marker size. It is seen from \fref{nenme} that both models yield a
nearly linear dependence. Only near threshold $N_e$ rises slowly for light
primaries, namely protons. The number of muons is expected to be a good
estimator for the primary energy, since, irrespective of the individual shower
development, the most abundant secondaries of the interactions are pions, for
which the charged species decay to muons and arrive to a large extent at the
Earth's surface.  This behavior is illustrated in \fref{nenme} (\rright). The
difference in energy for protons and iron nuclei for a fixed number of muons
amounts to about 25\% only for both models.

\section{Results}
\subsection{Primary energy correlations}

The number of electrons and muons registered at ground level as function of
energy for the interaction models EPOS~1.61 and QGSJET~01 is depicted in
\fref{nenme}.  For protons differences between the predictions of the two
models can be recognized. These differences are less pronounced for iron
induced showers.  The number of electrons is slightly lower for a fixed energy
for the model EPOS and the number of muons is larger for this model as
compared to QGSJET.

\begin{figure}
 \includegraphics[width=0.49\textwidth]{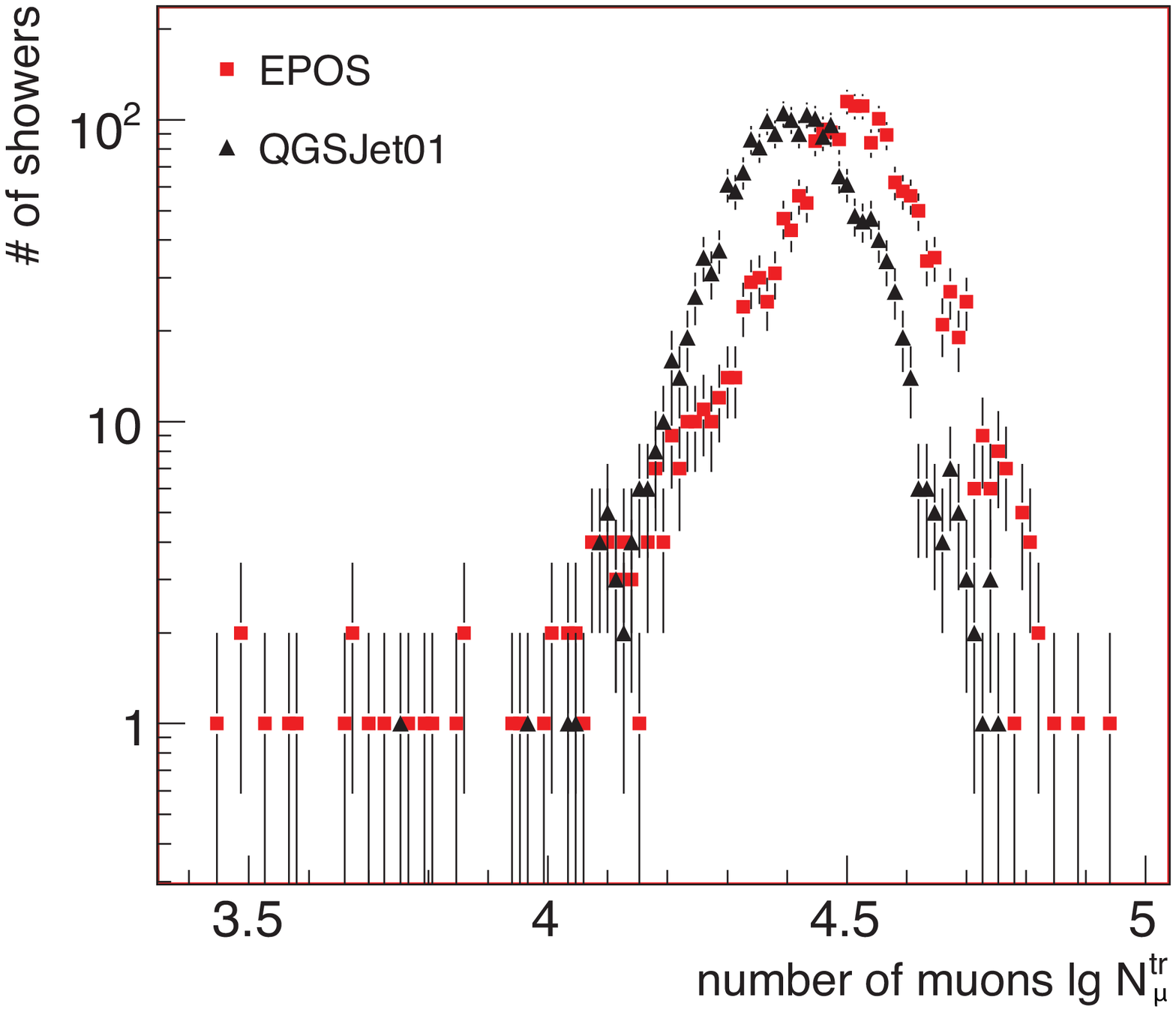}\hspace*{\fill}
 \includegraphics[width=0.49\textwidth]{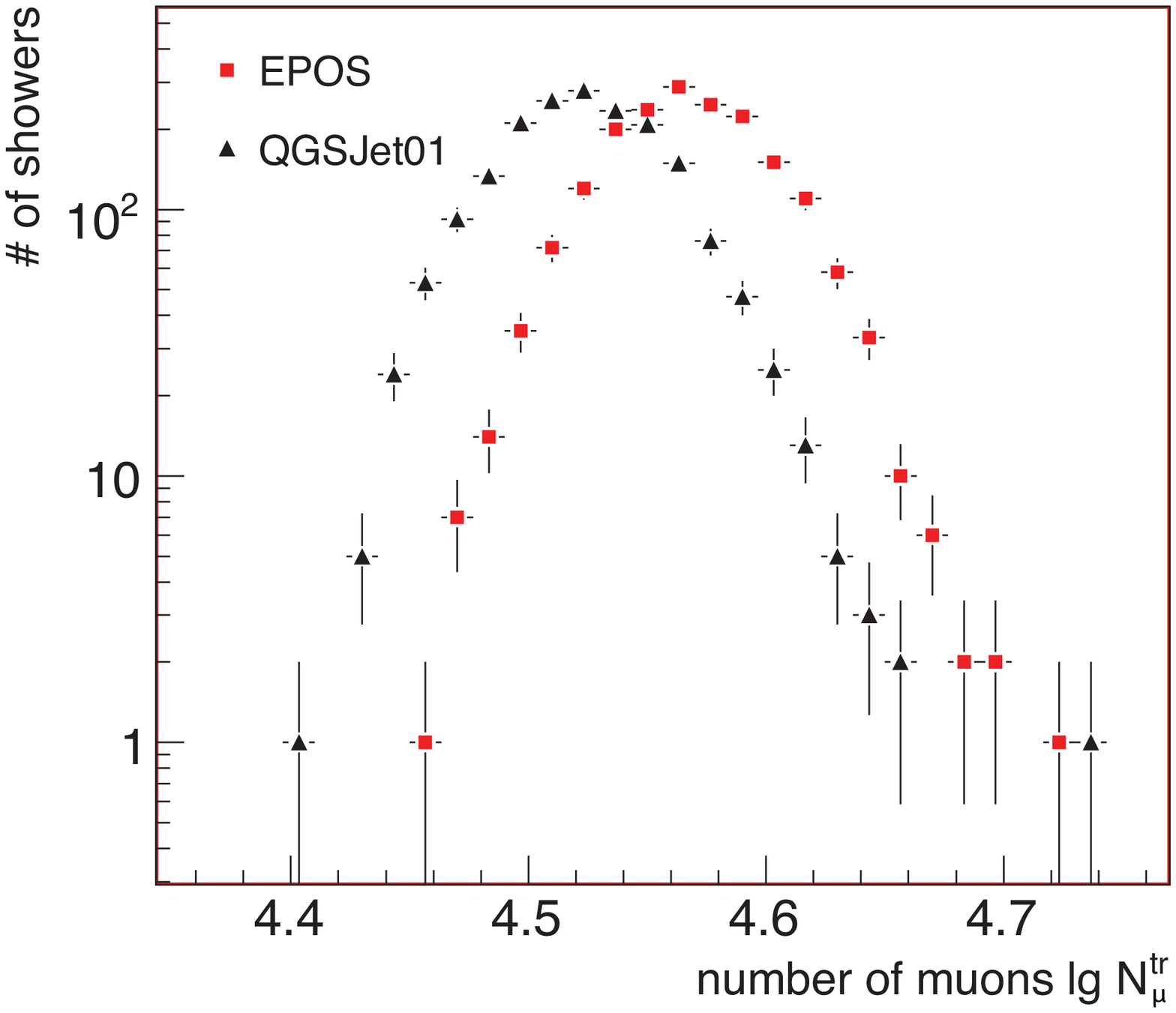}
 \includegraphics[width=0.49\textwidth]{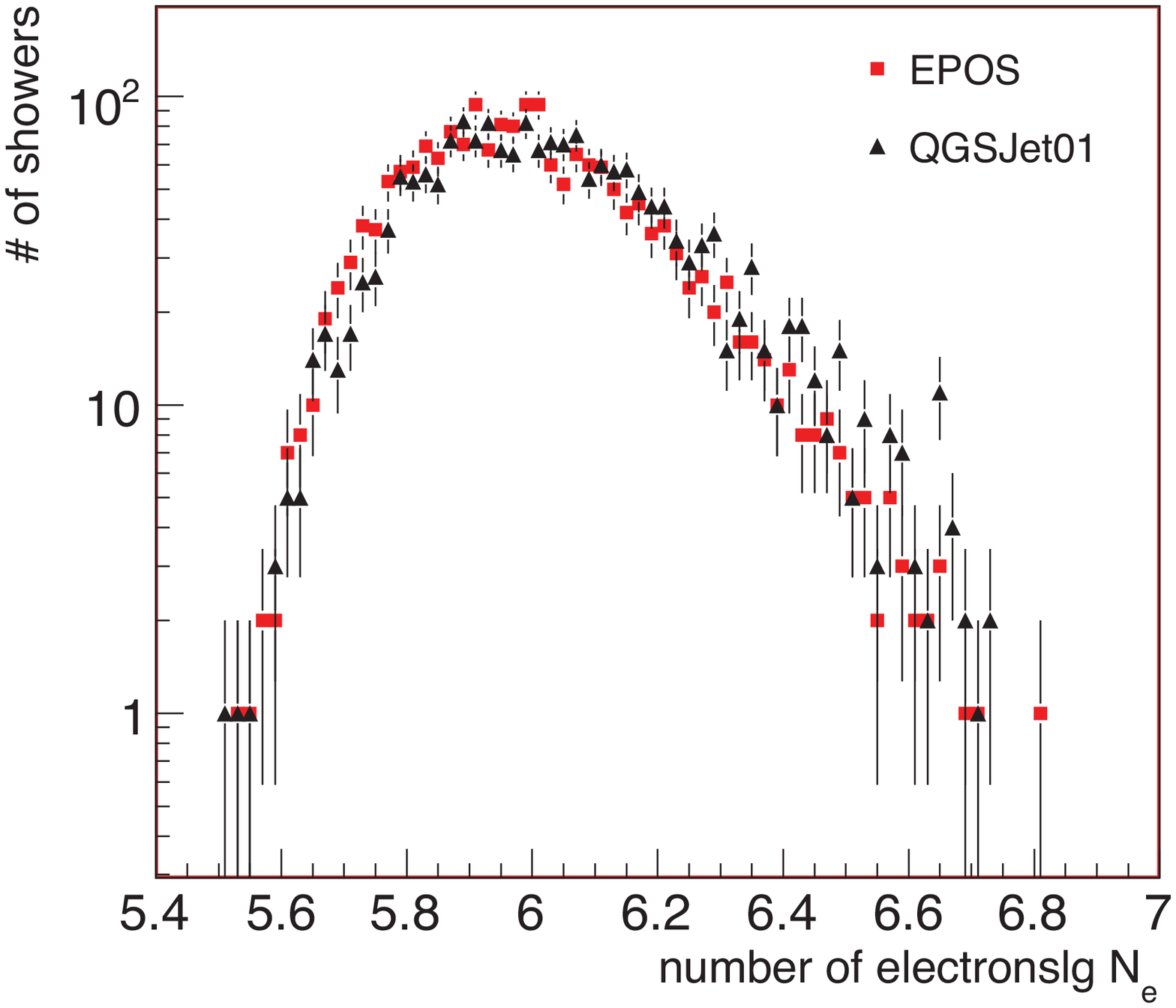}\hspace*{\fill}
 \includegraphics[width=0.49\textwidth]{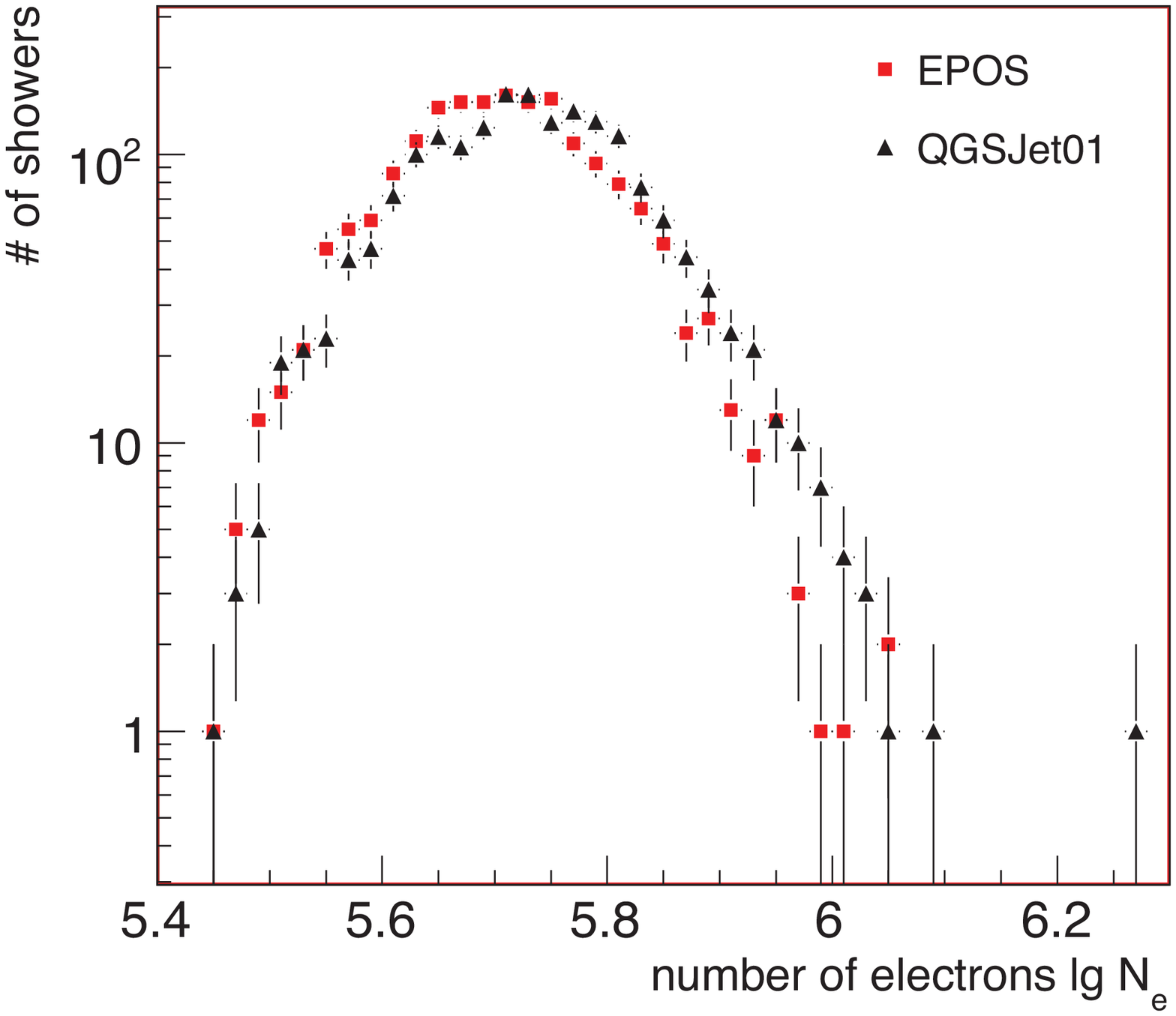}
 \caption{Predictions of two interaction models for the number of registered
	  muons (\ttop) and electrons (\bbottom) at ground for primary protons
          (\lleft) and iron nuclei (\rright) with an energy of $10^7$~GeV.}
 \label{qgs2}	  
\end{figure}

While mean values are shown in \fref{nenme}, the underlying distributions are
given in \fref{qgs2}. The figure displays the number of muons (\ttop) and
electrons (\bbottom) expected for showers with an energy of $10^7$~GeV for
primary protons (\lleft) and iron nuclei (\rright). Results for EPOS are
compared to predictions of the model QGSJET~01.  For both primary particle
species EPOS yields clearly more muons at observation level as compared to
QGSJET, while the shapes of the distributions are very similar. The
corresponding distributions for the number of electrons observed at ground
level are very similar for both models. They agree well in shape for both,
primary protons and iron nuclei. But the positions of the maxima are slightly
shifted.

\begin{figure}
 \includegraphics[width=0.49\textwidth]{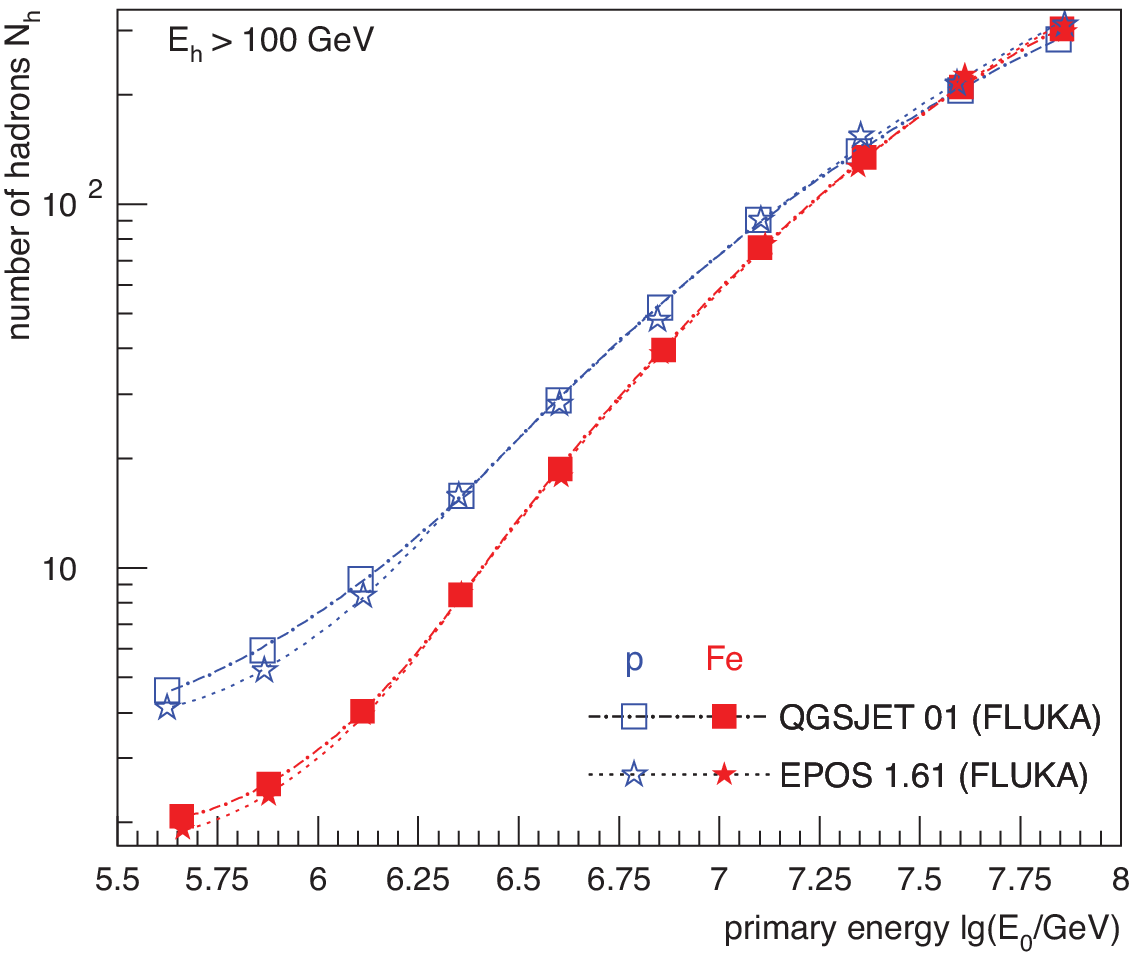}\hspace*{\fill}
 \includegraphics[width=0.49\textwidth]{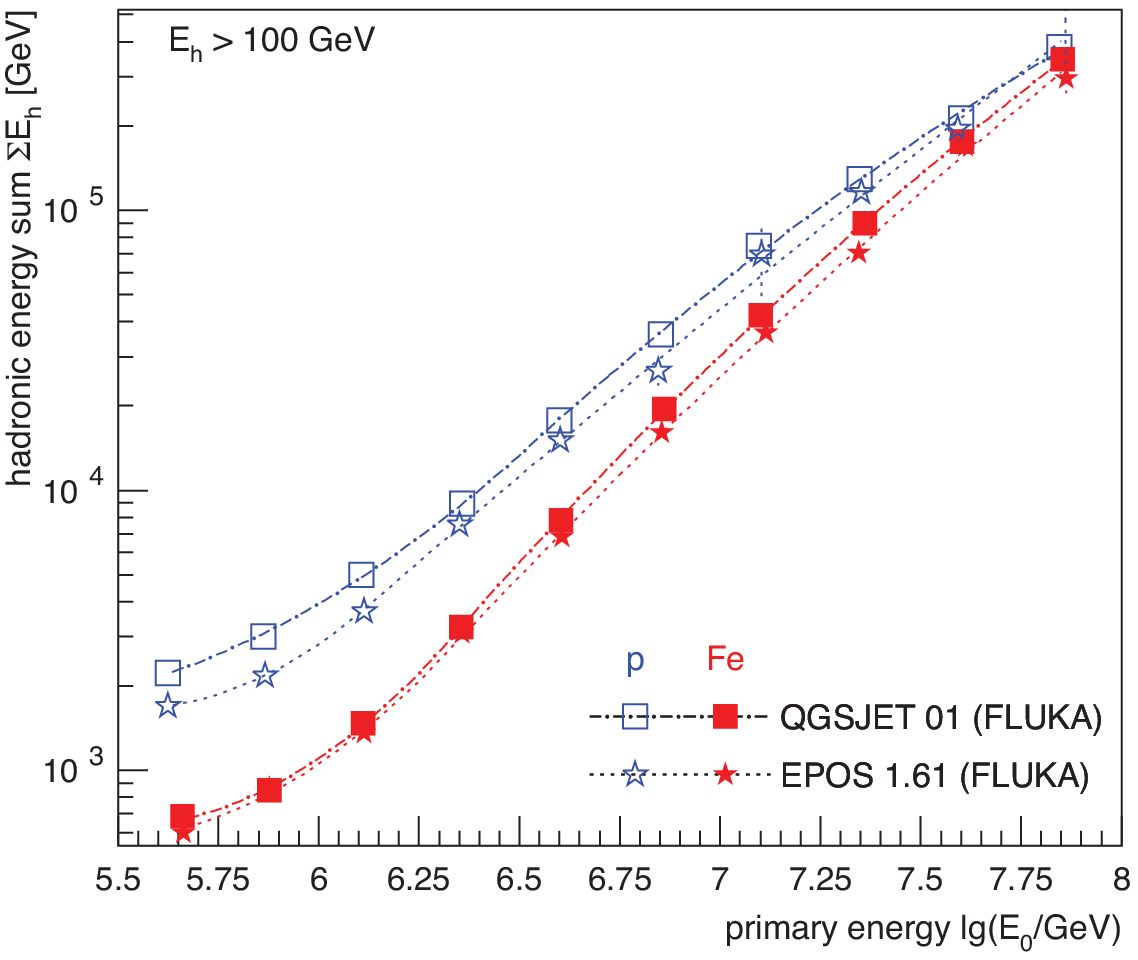}
 \caption{Number of hadrons (\lleft) and hadronic energy sum (\rright) as
	  function of shower energy for two interaction models and two primary
          particle species.}
 \label{nhehe}
\end{figure}

\begin{figure}
 \includegraphics[width=0.49\textwidth]{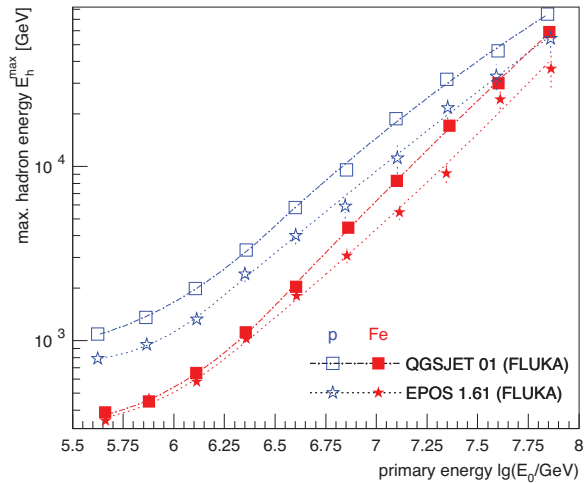}\hspace*{\fill}
 \begin{minipage}[b]{0.49\textwidth}
 \caption{Energy of the most energetic hadron in a shower as
	  function of shower energy for two interaction models and two primary
          particle species.}
 \label{mhe}
 \end{minipage}
\end{figure}

The relation of observed hadronic observables as function of shower energy are
presented in \ffref{nhehe} and \ref{mhe}. They show the number of hadrons
$N_h$, the hadronic energy sum $\sum E_h$, and the energy of the highest energy
hadron observed at ground $E_h^{max}$, respectively.  The number of hadrons for
a given energy predicted by both, EPOS and QGSJET~01 are very similar. But
there is a significant difference in the hadronic energy transported to
observation level. EPOS yields about 25\% smaller values for $\sum E_h$ as
compared to QGSJET~01 at the same shower energy. The effect has a similar
magnitude for both primary species. An even bigger difference is observed for
the value of the highest-energy hadron registered at ground level. The maximum
energies are reduced by up to 50\% to 60\% for EPOS compared to QGSJET at the
same shower energy.  Again, the effect is similarly strong for both primary
species.

\subsection{Electron -- muon correlations}

\begin{figure}
 \includegraphics[width=0.49\textwidth]{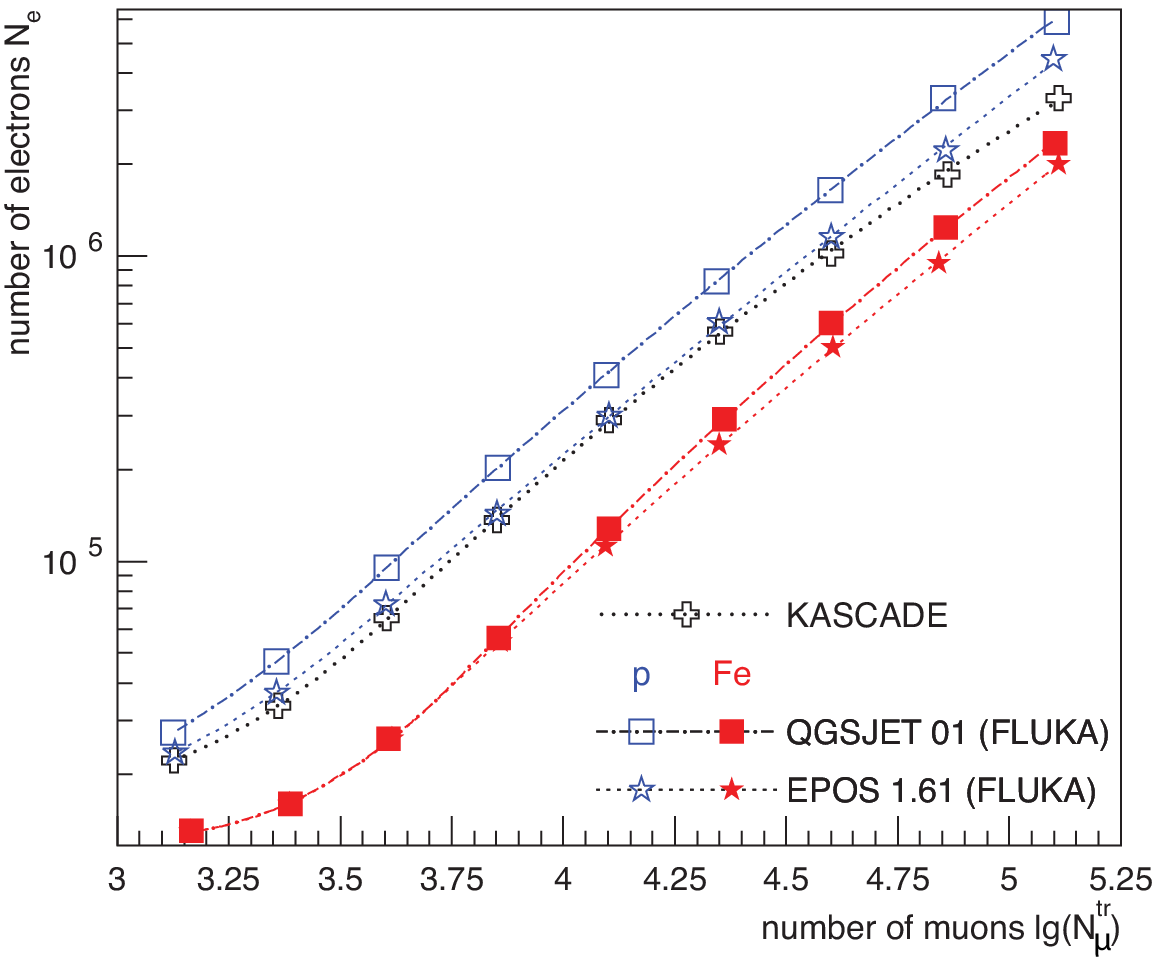}\hspace*{\fill}
 \includegraphics[width=0.49\textwidth]{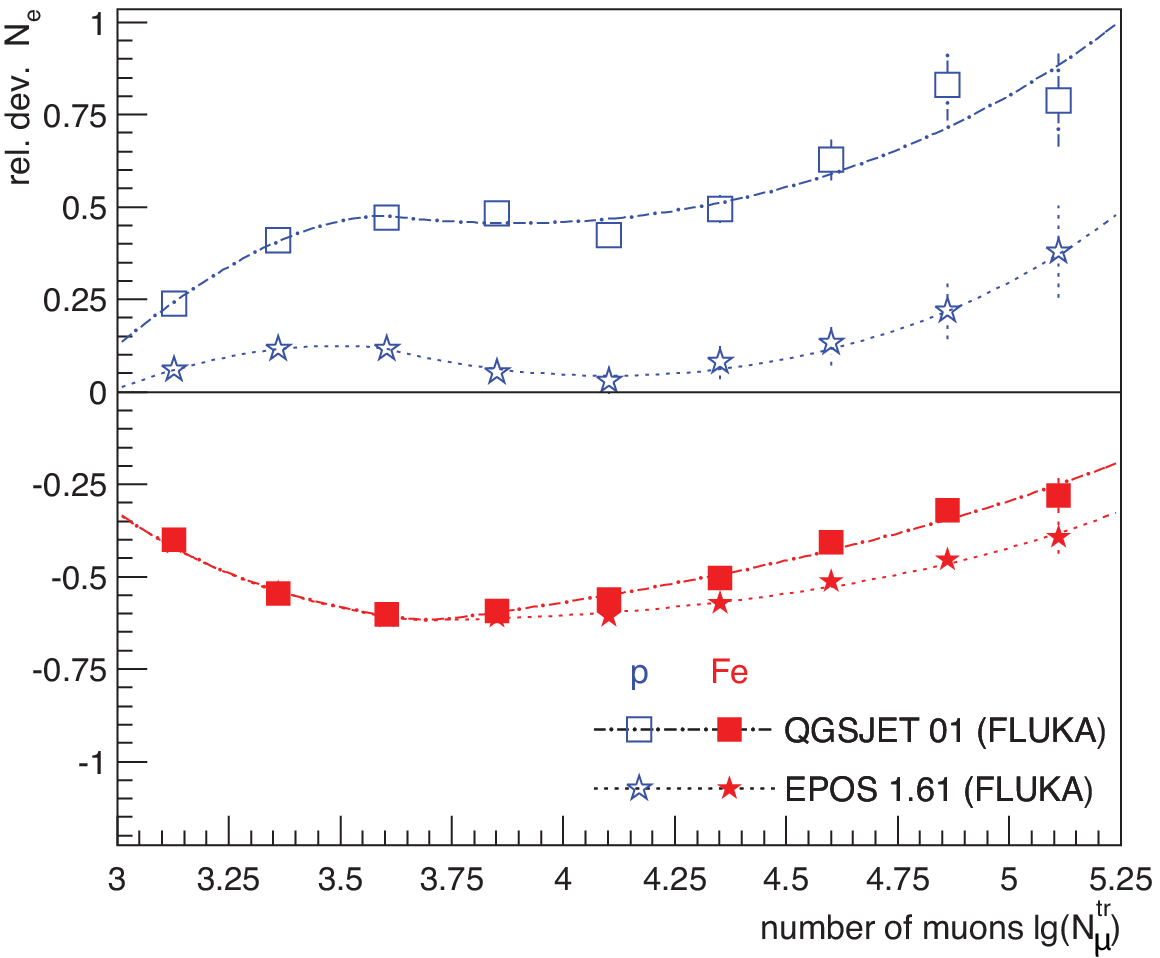}
 \caption{Number of electrons as function of muons for model predictions
	  compared to KASCADE measurements. Absolute values (\lleft) and
	  relative values $(N_e^{sim}-N_e^{meas})/N_e^{meas}$ (\rright).
	  Predictions for two interaction models and two primary particle
          species are shown.} 
 \label{nenm}	  
\end{figure}

Turning our attention towards observable quantities, among the most interesting
ones is the effect of the different models on the number of electrons and muons
at ground level. They are used to reconstruct energy and mass of the shower
inducing particles, e.g.\ by applying an unfolding algorithm
\cite{ulrichapp,ulrichisvhecri}.

The average number of electrons as function of the number of muons is displayed
in \fref{nenm} (\lleft) for the two models. Predictions for primary protons and
iron nuclei are compared to measured values. To emphasize the differences
between the model predictions, the same data are plotted on the right hand
panel in a different manner. The model predictions are shown relative to the
measured values, i.e.\ the quantity $(N_e^{sim}-N_e^{meas})/N_e^{meas}$ is
presented.
For a given muon number EPOS clearly yield less electrons ($\approx 40\%$) for
proton induced cascades and significantly lesser electrons for iron showers at
high energies, i.e.\ large muon numbers.
This can be understood taking \fref{nenme} into account. The differences seen
there for given primary energies translate into the significant discrepancies
between the two models seen in the electron-muon correlation (\fref{nenm}).

\begin{figure}
 \includegraphics[width=0.49\textwidth]{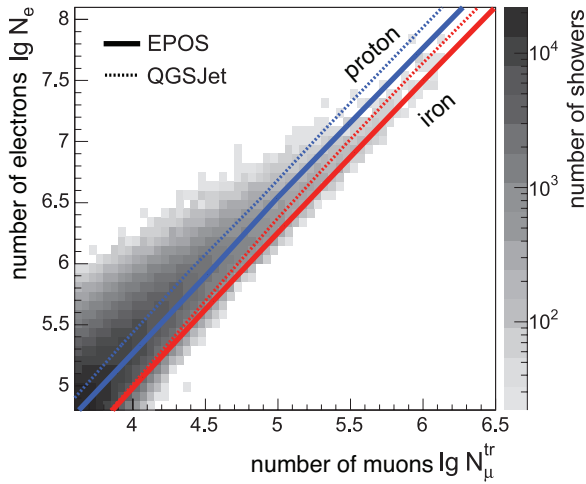}\hspace*{\fill}
 \begin{minipage}[b]{0.49\textwidth}
 \caption{Number of electrons as function of muons ($N_e-N_\mu$ plane).
 The measured two dimensional shower size distribution (grey shaded area) is
 compared to most probable values as predicted by two interaction models for
 two primary species.}
 \label{nenmplane}	  
 \end{minipage}
\end{figure}

To estimate the effects on the unfolding procedure it is useful to have a look
at the $N_e - N_\mu$ plane, see \fref{nenmplane}. The figure represents the
measured two dimensional shower size spectrum (grey coded area).  The lines
correspond to most probable values for primary protons and iron nuclei as
predicted by the interaction models EPOS and QGSJET.  It can clearly be
recognized that the lines for EPOS are shifted towards the lower right corner
of the figure with respect to QGSJET. This implies, if EPOS predictions are
used to derive the mass of the primary particles from the observed data a 
dominantly light mass composition is obtained.  

To check this effect, the energy spectra for five groups of elements (as in
\cite{ulrichapp}) have been unfolded from the measurements, based on EPOS
predictions. In this exercise in the energy range between $10^{6}$ and
$10^{7}$~GeV a very high flux of protons is obtained and the flux of heavy
particles (iron group) is strongly suppressed. If one extrapolates direct
measurements to high energies, such a behavior seems to be extremely
unrealistic.  This study illustrates that it would be very useful to measure
the energy spectra of individual elements directly up to the knee region.  Such
data would be very helpful to verify the interaction codes utilized in air
shower simulations.

\subsection{Hadron -- muon correlations}

\begin{figure}
 \includegraphics[width=0.49\textwidth]{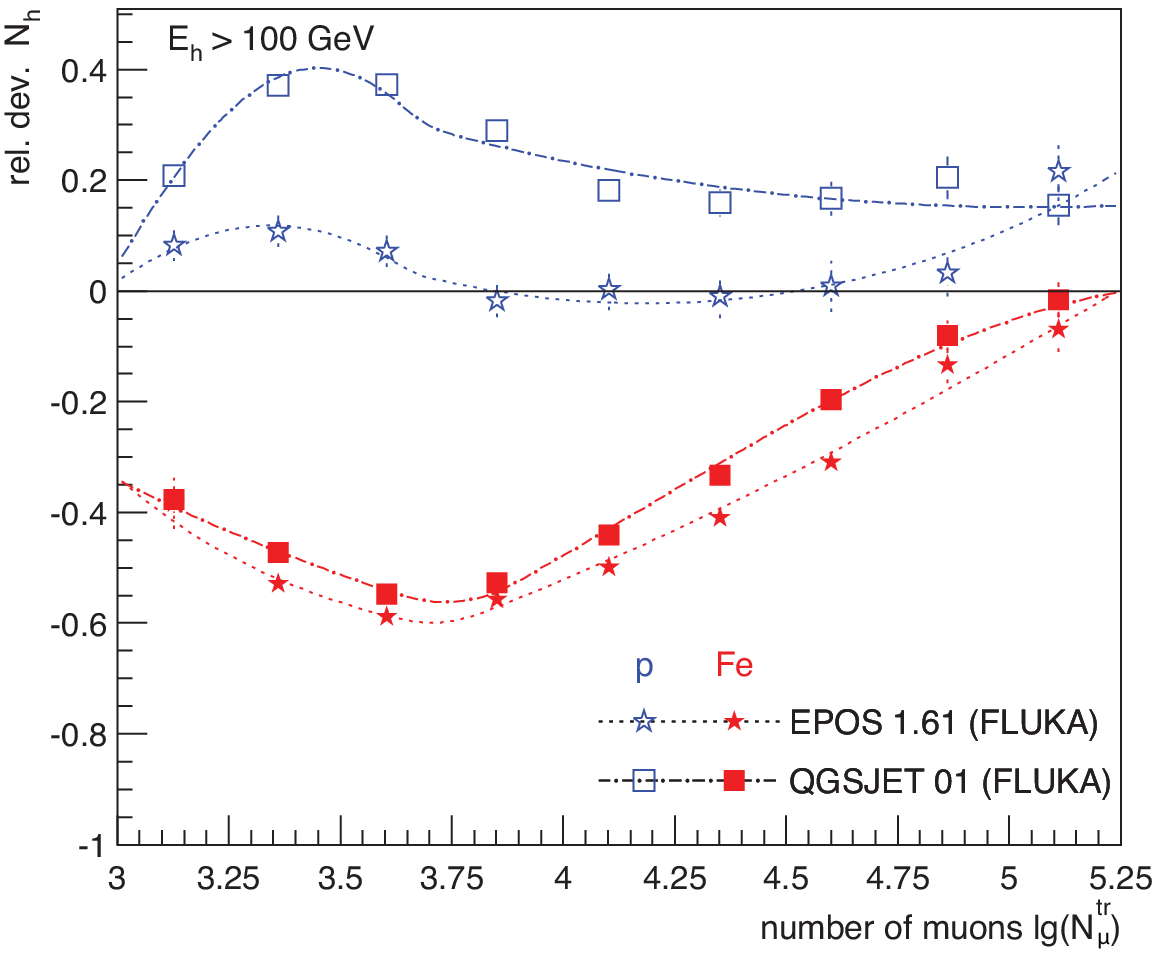}\hspace*{\fill}
 \includegraphics[width=0.49\textwidth]{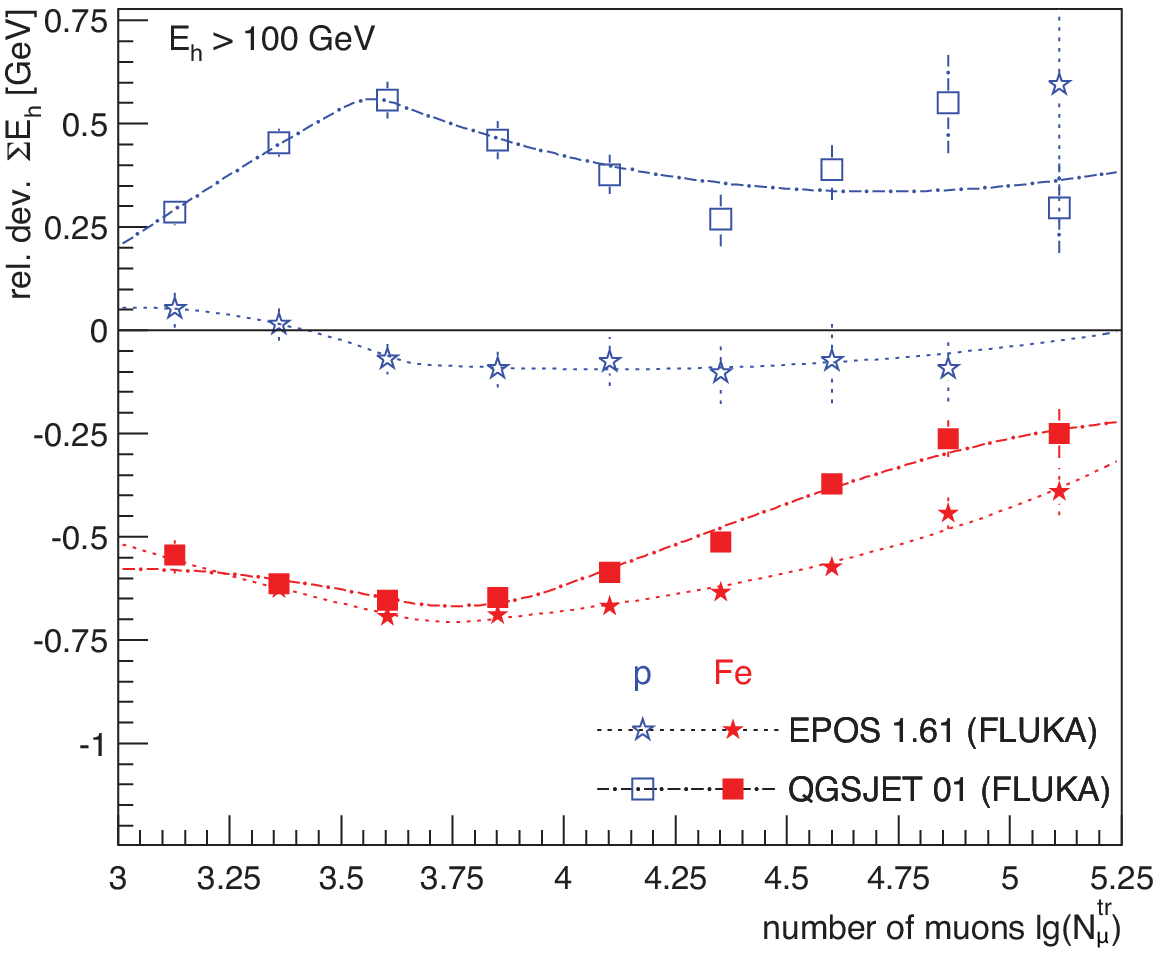}
 \caption{Number of hadrons observed (\lleft) and reconstructed 
 hadronic energy sum (\rright) as function of the registered number of muons
 for proton and iron induced showers.  The predictions of two interaction models
 are shown relative to the measured values.}
 \label{nhehm}	  
\end{figure}

The differences already seen in \fref{nhehe} are not directly accessible in
measurements, since the energy of the primary particle can not be inferred
directly. To check the validity of interaction models it is therefore suitable
to plot observable quantities against each other such as e.g.\ the number of
registered hadrons or the observed hadronic energy at ground level as function
of the number of muons as depicted in \fref{nhehm}.  Again, the model
predictions are plotted relative to the values measured by KASCADE, i.e.\ the
quantity $(x_{sim}-x_{meas})/x_{meas}$ is presented.  $x$ represents $N_h$ or
$\sum E_h$ for two interaction models and two primary particle species.  In
particular for primary protons for a given muon number EPOS yields
significantly less hadrons and delivers less hadronic energy to the observation
level.  It is generally assumed that in the energy range of interest the
average mass composition of cosmic rays is between protons and iron. Thus, in
\fref{nhehm} (as in \fref{nenm}, right) the zero line should be ``bracketed" by
the predictions for proton and iron induced showers, as it is the case for the
model QGSJET. On the other hand, it can be recognized that for EPOS at high
muon numbers (corresponding to energies around $10^{7}$~GeV) the hadronic
energy sums of both, proton and iron induced showers are smaller than the
experimental data. The systematic uncertainty of the hadronic energy sum
amounts to about 15\%. Within this uncertainty the data are compatible with the
EPOS predictions, assuming that all cosmic rays are protons only.  However, at
energies around $10^6$ to $10^7$~GeV this is not realistic. This implies that
the EPOS predictions are not compatible with the data.

The behavior observed for the maximum hadron energy $E_h^{max}$ registered at
observation level is very similar to the situation depicted in \fref{nhehm}
(\rright) for the hadronic energy sum. At high muon numbers EPOS yields values
for $E_h^{max}$ which are clearly below the measurements for both primary
species (protons and iron nuclei) --- an unrealistic scenario.

\begin{figure}
 \includegraphics[width=0.49\textwidth]{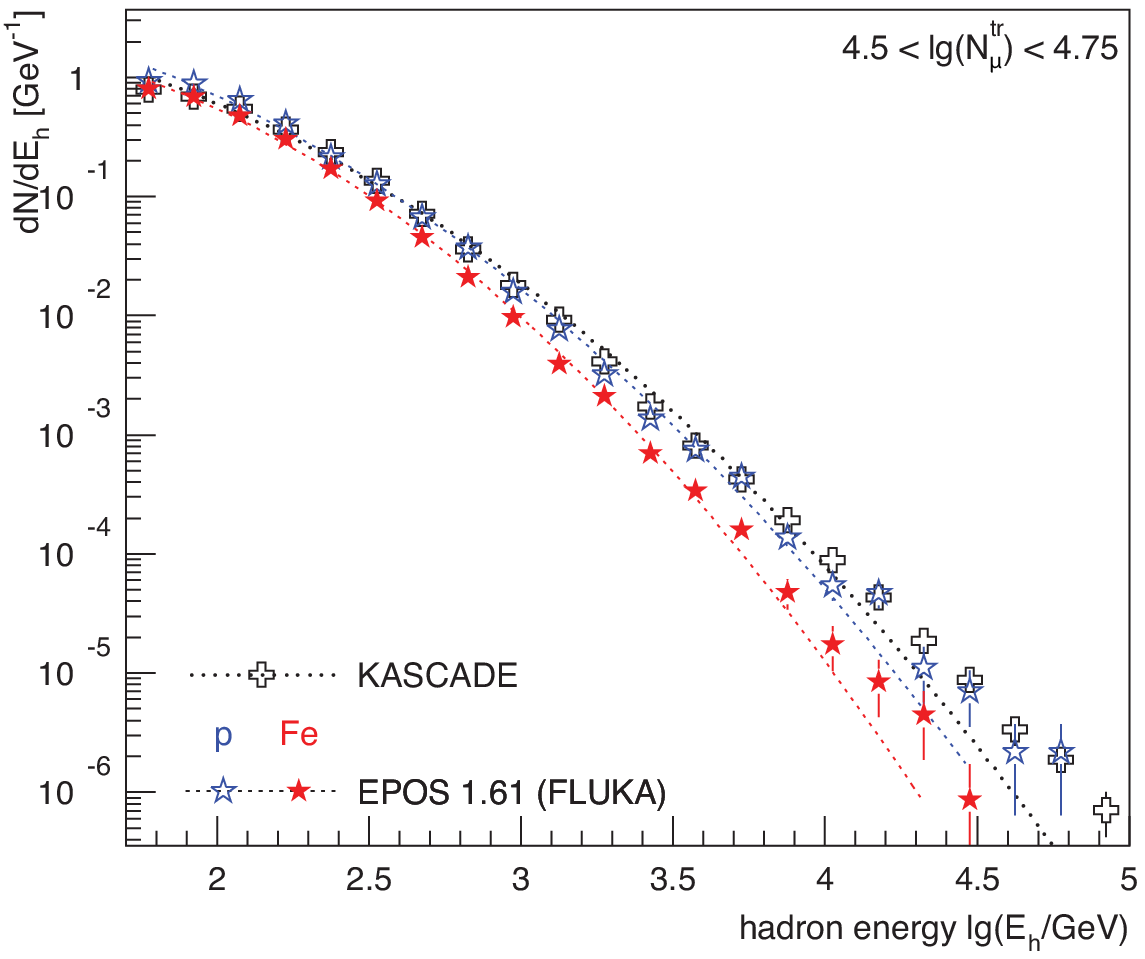}\hspace*{\fill}
 \includegraphics[width=0.49\textwidth]{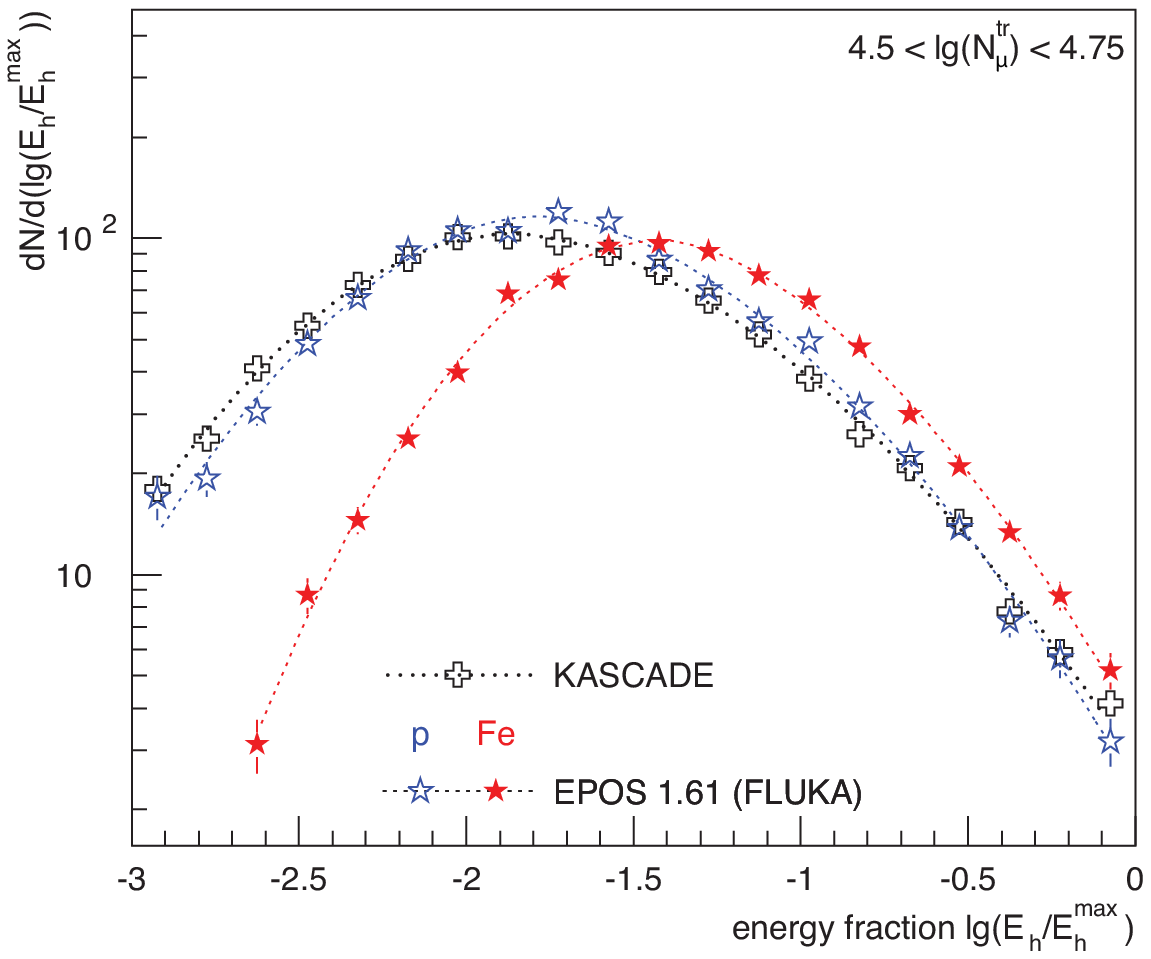}
 \caption{Energy spectrum of reconstructed hadrons (\lleft) and fraction of the
 reconstructed hadron energy to the maximum hadron energy (\rright) for a given
 muon number interval. EPOS predictions for proton and iron induced cascades are
 compared to measured values.}
 \label{hspec}	  
\end{figure}

Another way to shed light on the interaction models is to investigate the
energy spectra of hadrons for a given muon number interval, i.e.\ an
approximately fixed primary energy.  Hadron energy spectra as predicted by EPOS
are compared to measured values in \fref{hspec} (\lleft). 
\footnote{Corresponding distributions for the interaction model QGSJET have
been published previously \cite{wwtestjpg}. The predictions are compatible with
the measured data.}
The muon number interval corresponds to primary energies around
$2\cdot10^7$~GeV. At high hadron energies EPOS underestimates the observed
flux. The predictions for both primary species are below the measured values.
This observation is compatible with the above findings (\ffref{nhehe} and
\ref{mhe}), namely the relatively low hadronic energy sum and the relatively
small maximum hadron energy.

Distributions of the ratio of the energy of each reconstructed hadron to the
maximum hadron energy in each shower $E_h/E_h^{max}$ are plotted in
\fref{hspec} (\rright) for the same muon interval as above. Again, EPOS
predictions for two particle species are compared to measured data.  As for the
other observables discussed, the measurements should be ``bracketed" by the
predictions for proton and iron induced showers. However, the EPOS predictions
exhibit a clearly different behavior. For most $E_h/E_h^{max}$ ratios the
measured values are outside the proton-iron range given by the model.

The investigations of the energy spectra confirm the above findings, that EPOS
predictions are not compatible with KASCADE data.

\subsection{Hadron -- hadron correlations}

\begin{figure}
 \includegraphics[width=0.49\textwidth]{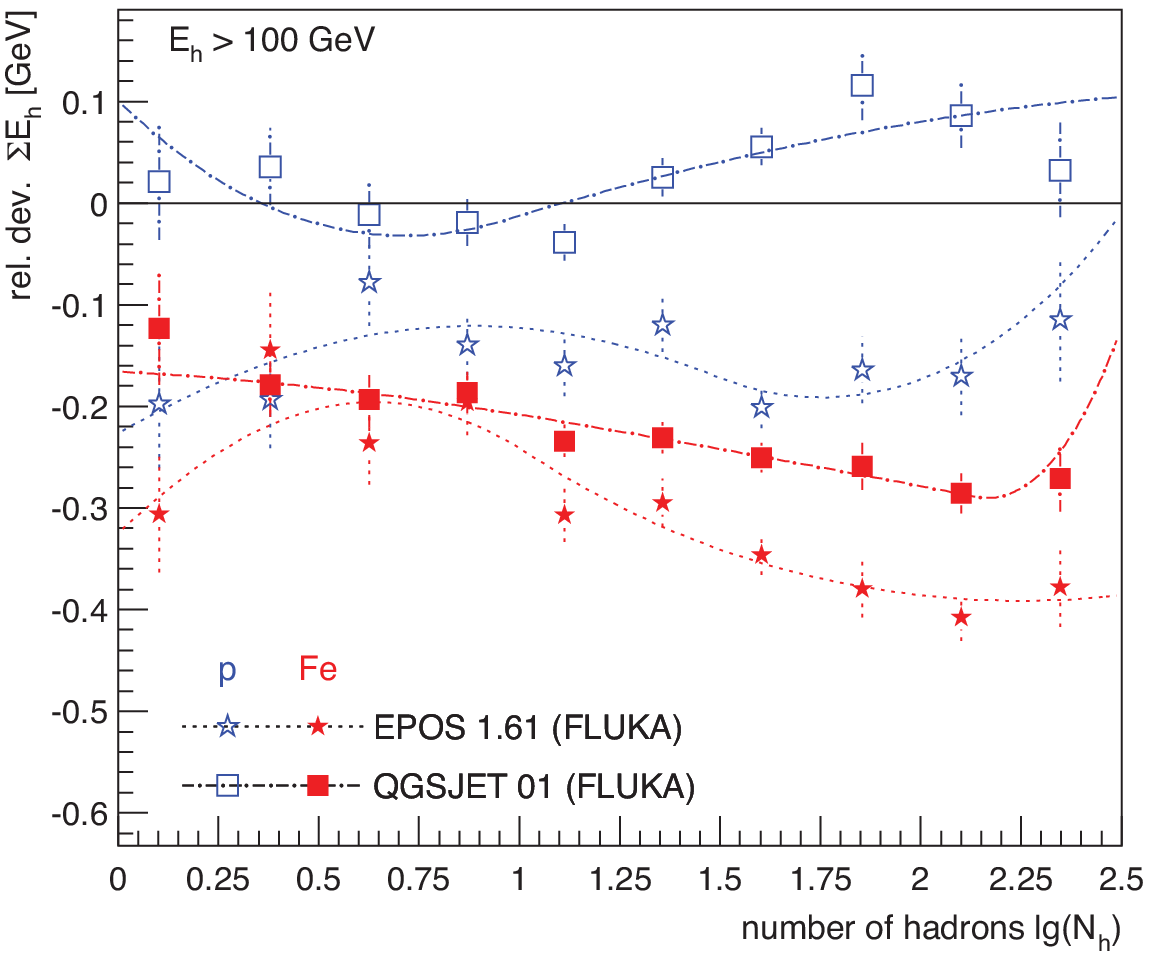}\hspace*{\fill}
 \includegraphics[width=0.49\textwidth]{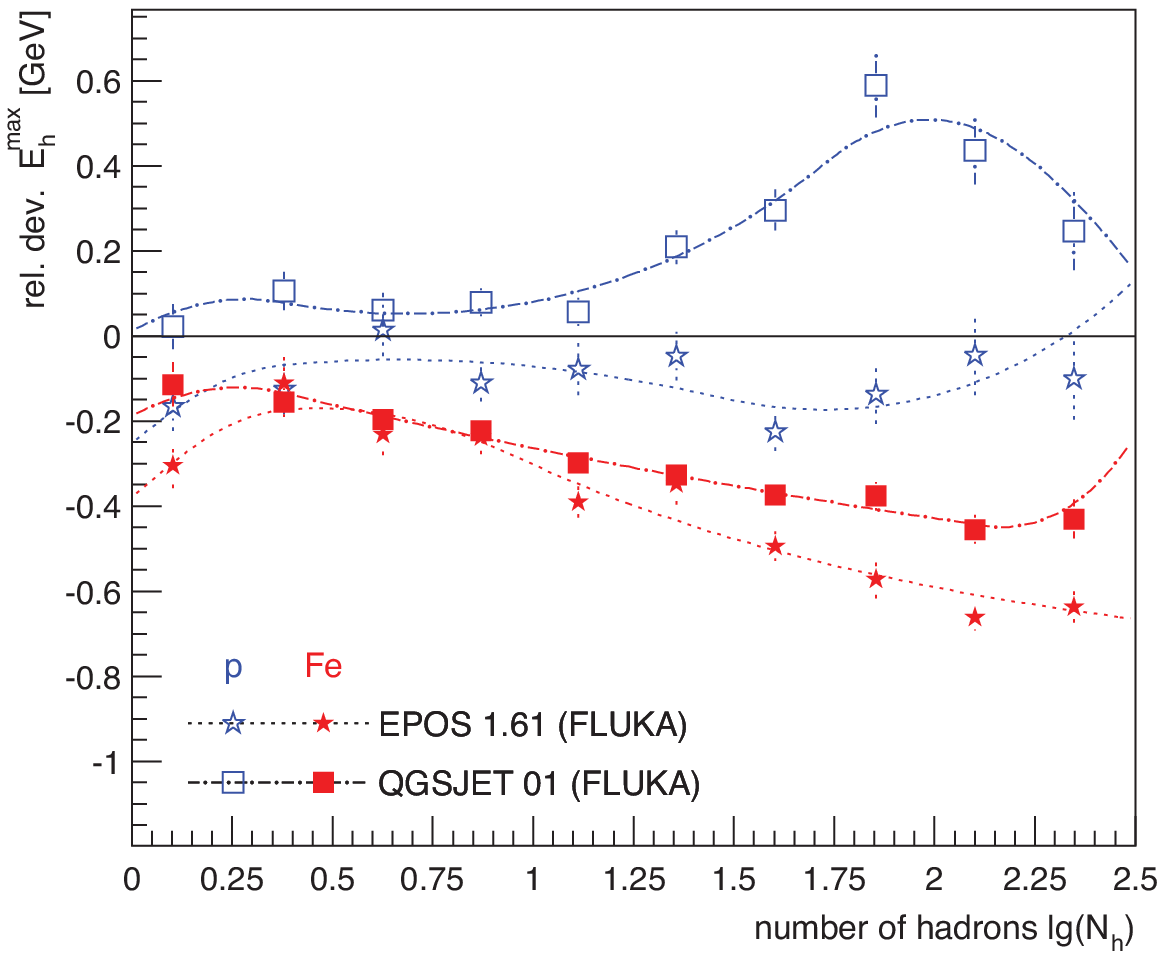}
 \caption{Relative hadronic energy sum $(\sum E_h^{sim}-\sum E_h^{meas})/\sum
  E_h^{meas}$ (\lleft) and relative maximum hadron energy (\rright) as function
 of the reconstructed number of hadrons for two interaction models and two
 primary particle species.}	  
 \label{hcorr}
\end{figure}

In the previous discussions it has been seen already that EPOS delivers less
energy in form of hadrons to ground level as compared to QGSJET~01. Therefore,
it is an interesting exercise to investigate also the correlations of the
purely hadronic observables with each other.  Examples of such correlations are
presented in \fref{hcorr}, depicting the hadronic energy sum (\lleft) and the
maximum hadron energy per shower (\rright).  The predicted values are again
plotted relative to the measured quantities to visually magnify the differences
between the model predictions.  In the figure the quantities are plotted as
function of the number of hadrons $N_h$.  Due to the steeply falling energy
spectrum and the $N_h-E_0$ correlation (see \fref{nhehe}) a sampling of the
data in $N_h$ intervals yields an enrichment of light particles. Therefore,
the data are expected to look very ``proton like".  Indeed, for QGSJET the
proton predictions are very close to the ``zero line", i.e.\ to the KASCADE
measurements. It should also be mentioned that (within the error bars) the
QGSJET predictions ``bracket" the measured values.  In contrast, the EPOS
predictions for both primary species are below zero for both observables shown
in the figure.  
The EPOS predictions for protons are at the lower bound of the 15\% systematic
uncertainty for the hadronic energy sum. Thus, they are barely compatible with
the data. However, it should be stressed that the QGSJET predictions for
protons really are at values around zero as expected. This indicates that the
systematic effects might be smaller than estimated and the EPOS predictions are
not compatible with the measurements.
From all observables investigated the hadron-hadron correlations exhibit the
strongest incompatibility between the EPOS predictions and the KASCADE data.

\section{Summary and conclusions}

Predictions of air shower simulations using the CORSIKA code with the hadronic
interaction models EPOS~1.61 and QGSJET~01 have been compared to measurements of
the KASCADE experiment. Various observables of the electromagnetic, muonic, and
hadronic component have been investigated and the correlations between them
have been analyzed. They have been used to check the compatibility of the EPOS
predictions with the KASCADE measurements.

The findings can be summarized as follows.
The investigations of the hadronic observables exhibit that EPOS does not
deliver enough hadronic energy to the observation level and the energy per
hadron seems to be too small.  In the $N_e-N_\mu$ plane the EPOS showers are
shifted to lower electron and higher muon numbers relative to QGSJET~01. When
the mass composition of cosmic rays is derived from measured values this effect
leads to a relatively light mass composition.
In summary, there is a significant discrepancy between the EPOS (version 1.61)
predictions and the KASCADE data. The EPOS predictions are not compatible with
the measurements.

\begin{figure}
 \includegraphics[width=0.49\textwidth]{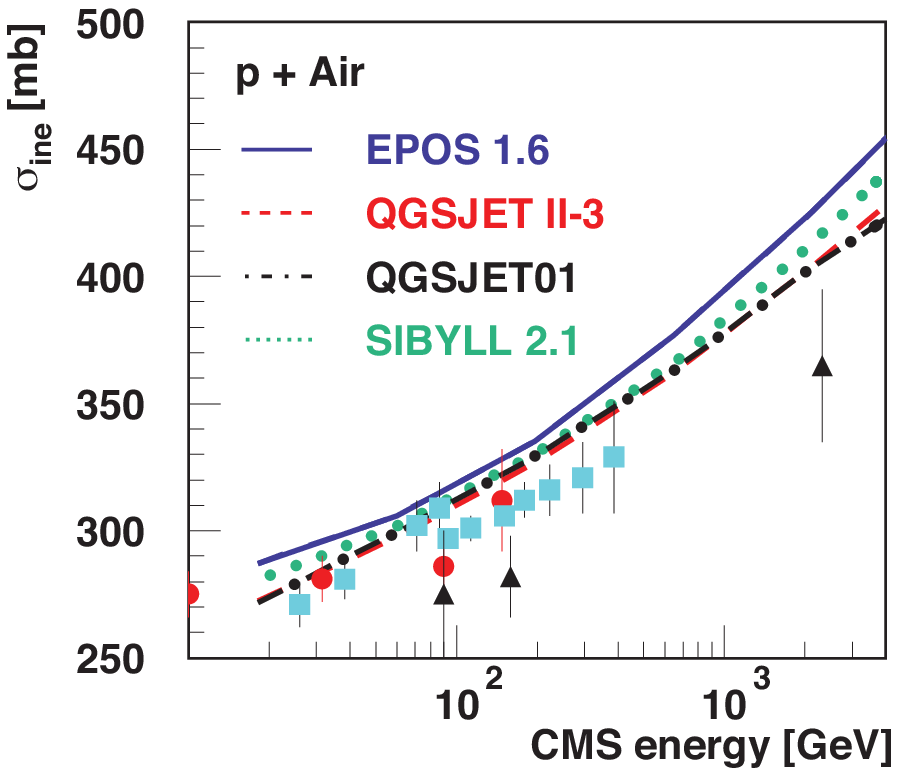}\hspace*{\fill}
 \includegraphics[width=0.49\textwidth]{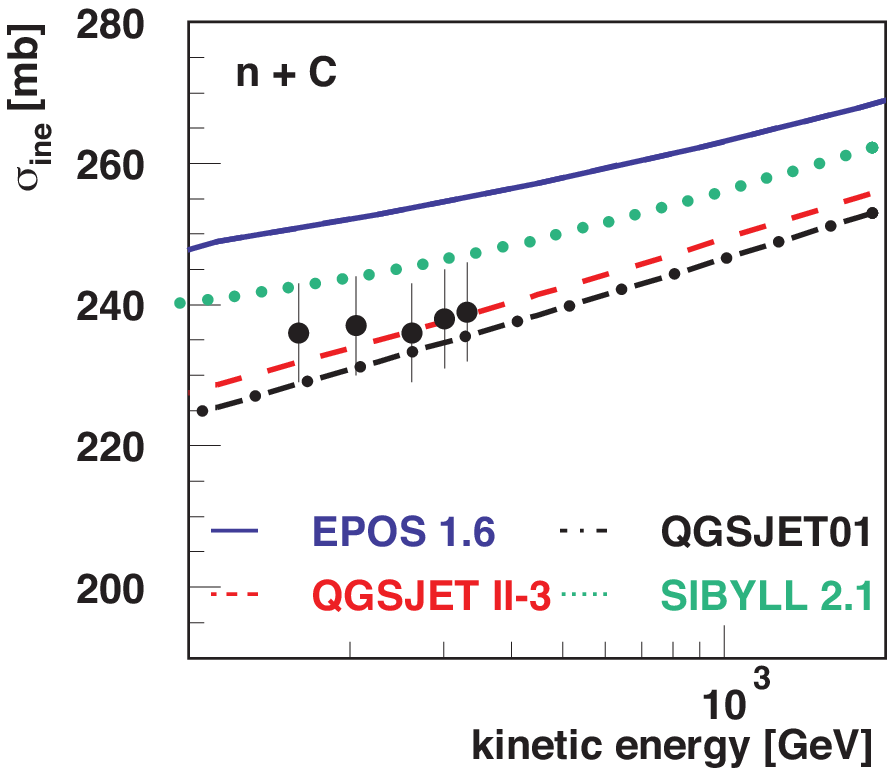}
 \caption{Inelastic cross sections for proton-air (\lleft) and neutron-carbon
  (\rright) collisions as predicted by various interaction models. The symbols
 represent experimental data, left: KASCADE prototype calorimeter (dots)
 \cite{mielkesh}, Yodh et al. (squares) \cite{yodh83}, ARGO-YBJ (triangles
 $\sim100$~GeV) \cite{argo-merida} and EAS-TOP (triangle $\sim 2000$~GeV)
 \cite{eastop-merida}; 
 right: Roberts et al.\ \cite{roberts1979}.} 
 \label{crosssect}
\end{figure}

Most likely the incompatibility of the EPOS predictions with the KASCADE
measurements is caused by too high inelastic cross sections for hadronic
interactions implemented in the EPOS code. To illustrate this, the proton-air
and neutron-carbon cross sections as predicted by different models are
displayed in \fref{crosssect}. It can be recognized that the EPOS~1.61 values
mark the upper limit of the variations exhibited by the different models.
Already at moderate energies in the 100~GeV regime a clear difference between
the models is visible. In particular, the example of the neutron-carbon cross
section illustrates that even at energies accessible to todays accelerator
experiments, the models contain different descriptions of the inelastic
hadronic cross sections.
According to the authors of the EPOS code, a new version is in preparation with
lower cross sections. It is expected that the predictions of this version are
in better agreement with air shower data. Further studies shall be presented in
a follow-up publication.

The results presented also underline the importance of measuring hadronic
observables in air shower experiments. They provide the most sensitive
available means of investigating the properties of hadronic interactions at
very high energies and kinematical ranges to complement accelerator
experiments.

\section*{Acknowledgement}
The authors would like to thank the members of the
engineering and technical staff of the KASCADE-Grande
collaboration, who contributed to the success of the experiment.
The KASCADE-Grande experiment is supported
by the BMBF of Germany, the MIUR and INAF of Italy,
the Polish Ministry of Science and Higher Education, 
and the Romanian Ministry of Education and Research 
(grant CEEX 05-D11-79/2005).

\section*{References}

\end{document}